\documentclass[prl,aps,twocolumn,superscriptaddress,showpacs]{revtex4}
\usepackage{graphicx}

\begin{document}

\title{Magnetic Moment of the Fragmentation-Aligned $^{61}Fe(9/2 ^+)$ Isomer.}

\author{I.~Matea}\affiliation{GANIL, BP 55027, 14076 Caen Cedex 5, France}
\author{G.~Georgiev}\affiliation{GANIL, BP 55027, 14076 Caen Cedex 5, France}
\author{J.M.~Daugas}\affiliation{CEA/DIF/DPTA/PN, BP 12, 91680 Bruy\`eres
le Ch\^atel, France}
\author{M.~Hass}\affiliation{The Weizmann Institute, Rehovot, Israel}
\author{G.~Neyens}\affiliation{University of Leuven, IKS, Celestijnenlaan
200 D, 3001 Leuven, Belgium}
\author{R.~Astabatyan}\affiliation{FLNR-JINR,~Dubna,~Russia}
\author{L.T.~Baby}\affiliation{The Weizmann Institute, Rehovot, Israel}
\author{D.L.~Balabanski}\affiliation{Faculty of Physics, St. Kliment
Ohridski University of Sofia, 1164 Sofia, Bulgaria.}
\author{G.~B\'elier}\affiliation{CEA/DIF/DPTA/PN, BP 12, 91680 Bruy\`eres
le Ch\^atel, France}
\author{D.~Borremans}\affiliation{University of Leuven, IKS, Celestijnenlaan
200 D, 3001 Leuven, Belgium}
\author{F.~Chappert}\affiliation{CEA/DIF/DPTA/PN, BP 12, 91680 Bruy\`eres le
Ch\^atel, France}
\author{M.~Girod}\affiliation{CEA/DIF/DPTA/PN, BP 12, 91680 Bruy\`eres le
Ch\^atel, France}
\author{G.~Goldring}\affiliation{The Weizmann Institute, Rehovot, Israel}
\author{H.~Goutte}\affiliation{CEA/DIF/DPTA/PN, BP 12, 91680 Bruy\`eres le
Ch\^atel, France}
\author{P.~Himpe}\affiliation{University of Leuven, IKS, Celestijnenlaan
200 D, 3001 Leuven, Belgium}
\author{M.~Lewitowicz}\affiliation{GANIL, BP 55027, 14076 Caen Cedex 5, France}
\author{S.~Lukyanov}\affiliation{FLNR-JINR,~Dubna,~Russia}
\author{V.~M\'eot}\affiliation{CEA/DIF/DPTA/PN, BP 12, 91680 Bruy\`eres le
Ch\^atel, France}
\author{F.~de~Oliveira~Santos} \affiliation{GANIL, BP 55027, 14076 Caen
Cedex 5, France}
\author{Yu.E.~Penionzhkevich}\affiliation{FLNR-JINR,~Dubna,~Russia}
\author{M-.G.~Porquet}\affiliation{CSNSM,~Orsay,~France}
\author{O.~Roig}\affiliation{CEA/DIF/DPTA/PN, BP 12, 91680 Bruy\`eres le
Ch\^atel, France}
\author{M.~Sawicka}\affiliation{IFD, Warsaw University, Ho\.{z}a 69, 00681
Warsaw, Poland}

\date{\today}


\begin{abstract}

We report on the g factor measurement of the isomer in $^{61}Fe$
($E^{*}=861~keV$). The isomer was produced and spin-aligned via a
projectile-fragmentation reaction at intermediate energy, the Time
Dependent Perturbed Angular Distribution (TDPAD) method being used
for the measurement of the g factor. For the first time, due to
significant improvements of the experimental technique, an
appreciable residual alignment of the isomer has been observed,
allowing a precise determination of its g factor: $g=-0.229(2)$.
Comparison of the experimental g factor with shell-model and mean
field calculations confirms the $9/2^+$ spin and parity
assignments and suggests the onset of deformation due to the
intrusion of Nilsson orbitals emerging from the $\nu g_{9/2}$.
\end{abstract}


\maketitle

The measurement of electromagnetic (EM) moments has traditionally
played a central role in the critical evaluation of nuclear
structure models since they elucidate the single-particle nature
(magnetic moments) and the shape (quadrupole moments) of the
nuclear state under investigation. Their particular behavior
around nuclear shell closures, approaching Schmidt values for
magnetic moments and small values for quadrupole moments, makes
them good tools to investigate shell closure near and far from
$\beta$ stability.

One of the main restrictions in the study of EM moments is the
necessity to obtain oriented isomeric ensembles. In particular,
for the study of neutron rich nuclei, that are mostly produced via
fragmentation reactions at high or intermediary energies, the
mechanism to produce oriented states is not yet well understood.
The first observation of spin alignment of isomeric states in a
projectile-fragmentation reaction at an energy of 500 MeV/u was
reported by Schmidt-Ott {\it et al} \cite{schmidtott94}. At
projectile energies below 100 MeV/u, so called intermediate
energies, spin alignment of isomeric beams was reported recently
by Georgiev {\it et al} \cite{georgiev02} in the study of isomeric
states in the $^{68}Ni$ region. However, only a very small
residual alignment observed in the decay of the $^{67}Ni$ and the
$^{69}Cu$ isomers was reported.

In this letter, we report an important experimental achievement in
the study of EM moments of neutron rich isomers. For the first
time, high quality data could be obtained for an isomer in the
neutron rich $^{61}Fe$ nucleus, located near the controversial
N=40 subshell closure
\cite{bern82,oros00,broda95,leenhardt_epja02,langanke03,sor02_prl88}.

We focus here on the role played by the $\nu g_{9/2}$ orbital in
the low-energy level structure of nuclei near $N=40$ in the
particular case where this orbital manifests itself as an isomeric
state. The $^{61m}Fe$, with 35 neutrons, is one of the lightest
nuclei exhibiting such an isomeric state ($T_{1/2}=250(10)$ ns) at
a rather low excitation energy ($E^{*}=861$ keV). The $9/2^+$
tentatively assigned spin and parity is based on systematics
\cite{grzywacz98}. The measurement of the g factor of the isomeric
state in $^{61}Fe$ provides information about its structure and
can also confirm the suggested spin/parity.

\begin{figure}[t]
\includegraphics[scale=0.7]{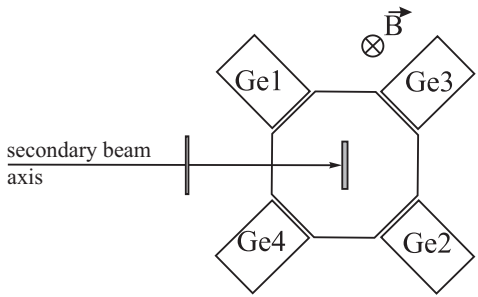}
  \caption{Schematic drawing of the TDPAD experimental set-up. The
beam passes through a 50 $\mu$m plastic scintillator before being
stopped in a 500 $\mu$m Cu foil.} \label{fig:setup}
\end{figure}

The nuclei of interest were produced following the fragmentation
of a 54.7 MeV/u $^{64}$Ni beam accelerated at the GANIL facility,
with a mean intensity of $7\cdot 10^{11}$ pps, impinging on a 97.6
$mg/cm^2$ thick $^{9}$Be target placed at the entrance of the LISE
spectrometer \cite{anne87}. In order to decrease the in-flight
decay of the isomer, the detection setup (Fig. \ref{fig:setup})
was positioned at the first focal plane of the LISE spectrometer
(time of flight $\approx 200$ ns). A 300~${\mu}m$ thick removable
silicon detector was used to optimize the selection of $^{61}Fe$
fragments by means of energy-loss vs time-of-flight
identification. Once the selection was performed, a 50 ${\mu}m$
thick plastic scintillator, placed in front of the catcher foil,
was used to provide the $t=0$ signal for the subsequent TDPAD
measurement. The choice of a thin plastic scintillator instead of
a silicon detector was made for two reasons. First, the fragments
are selected fully stripped with the LISE spectrometer and, in
order to preserve the initial alignment until the eventual stop in
a catcher foil, it is important to avoid as much as possible the
capture of electrons that can completely destroy the orientation
of the nuclear ensemble. The electron pick-up upon passing through
the scintillator is estimated to be below 2\% \cite{lise_prog},
much lower than that in a 300 ${\mu}m$ thick silicon detector
(60-70\%) \cite{georgiev02}. Secondly, the accepted secondary beam
rate (between 17 and 80 kHz for the present experiment) can be one
order of magnitude higher than with a silicon detector. As a
catcher for the reaction products we used an annealed high-purity
500 ${\mu}m$ thick Cu foil. Iron ions have the same
electronegativity and similar atomic radius as Cu atoms and hence
Cu, with its cubic lattice structure, is expected to provide a
perturbation-free environment for implanted Fe fragments. The Cu
foil was placed between the poles of an electromagnet that
provided a constant magnetic field \overrightarrow{B} in the
vertical direction.

The Larmor precession of the initially aligned isomeric spins in
the applied field of about 0.7 T, was monitored with four coaxial
Ge detectors placed in the horizontal plane around the Cu foil as
shown in Fig. \ref{fig:setup}. Time spectra were collected, having
as start the signal due to the ion passage through the plastic
scintillator and as stop the signal given by the detection of a
prompt or delayed $\gamma$-ray. To extract the precession pattern
out of the time spectra, data from detectors positioned at 90$^o$
with respect to each other was combined to generate the
$R(t)$-function:
\begin{eqnarray}
\label{fo:Rt} R(t)&=& \frac {I_{12}(\theta,t) - \epsilon
I_{34}(\theta+\frac{\pi}{2},t)}{ I_{12}(\theta,t) +
\epsilon I_{34}(\theta+\frac{\pi}{2},t)} \nonumber\\
&\sim& A_2 B_2^0(t=0) cos(2(\omega_L t+\alpha-\theta)).
\end{eqnarray}

\noindent with $\overrightarrow{\omega}_{L} $=$ - \frac {g
\mu_N\overrightarrow{B}}{\hbar}$ and $\alpha $=$ - \frac
{\pi}{2}(1 - \frac{gA}{2Z})$ both depending on the g factor.
$I_{12}$ and $I_{34}$ are the summed intensities of detectors
placed at 180 degrees (see figure \ref{fig:setup}); $A_2$ is the
radiation parameter of the $\gamma$-ray transition; $B_2^0$ is the
second component of the orientation tensor describing the initial
orientation of the selected isomeric ensemble; $\theta$ equals
$\pi/4$; $\epsilon$ is the relative efficiency between the four
respective detectors in the present setup, and $\alpha$ is the
rotation angle of the aligned ensemble symmetry axis with respect
to the beam axis induced when passing through the two dipole
magnets of LISE spectrometer \cite{georgiev02}. The data
acquisition was validated on an event by event basis by the
coincidence between a heavy ion signal from the plastic
scintillator and a delayed $\gamma$, registered by one of the
germanium detectors, within a time window of 3 ${\mu}s$. In order
to diminish the number of accidental coincidences in this time
window, we have used a package suppresser that brought on target 1
out of 10 ion packages provided by the accelerator with a
frequency of 10.5 MHz. As a consequence, the detected alignment
increased by a factor of $\sim$ 3 compared to a measurement
without package suppression.

\begin{figure}[t]
\includegraphics[scale=0.54]{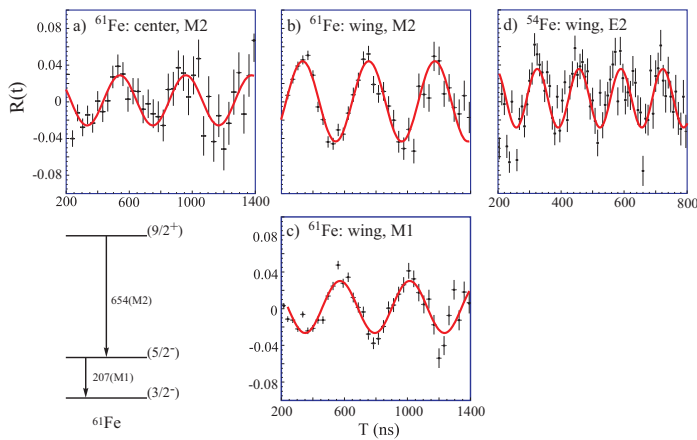}
  \caption{The R(t) functions for: a.-$^{61}Fe$(center, 654 keV);
b.-$^{61}Fe$(wing, 654 keV); c.-$^{61}Fe$(wing, 207 keV);
d.-$^{54}Fe$. The R(t) function for 207 keV is consistent with
that for the 654 keV but with opposite phase, as expected from the
assumed multipolarities of the corresponding $\gamma$
transitions.} \label{fig:rt_fits}
\end{figure}

Certain physical effects need to be taken into account when
deducing a precise value for the g factor from the observed Larmor
frequency. The distribution of the magnetic field over the beam
spot, the paramagnetic amplification of the applied magnetic field
and the Knight shift, all can induce minor modifications of the
Larmor frequency. In order to avoid systematic errors due to these
corrections and to validate the experimental setup and method, we
have measured the Larmor precession under identical conditions for
isomers in two Iron isotopes: the $I^{\pi}=10^{+}$, $E^*=6527$ keV
isomer in $^{54}Fe$ having a known g factor, $g(10^{+}) =
+0.7281(10)$ \cite{rafailovich83}, and the $I^{\pi}$=$(9/2^{+})$,
$E^{*}=861$ keV isomer in $^{61}$Fe decaying via a (M2) transition
of 654 keV in cascade with a (M1) transition of 207 keV to the
ground state (see fig. \ref{fig:rt_fits}). The effective value of
the magnetic field extracted from the R(t) function of $^{54m}Fe$
(eq. \ref{fo:Rt}) is 6800(40) Gs and it includes all the
corrections mentioned above.

$^{54m}Fe$, decaying by a stretched E2-cascade, served also as a
probe to measure the produced alignment of the isomeric ensembles
as a function of their angular momentum distribution
\cite{neyens03}. For the selection of the fragments in the wing of
the momentum distribution ($p_{fragment}<p_{projectile}$), the
amplitude of the R(t) function (fig. \ref{fig:rt_fits}) yields a
large negative alignment, $-12.5(9)\%$, the sign being in
agreement with the predictions of a kinematical fragmentation
model \cite{asahi91,daugas01}.

For $^{61m}Fe$, we have studied the alignment for two ensembles
selected, respectively, in the center and in the outer wing of the
longitudinal momentum distribution. The alignment of the isomeric
state, deduced from the amplitude of the R(t) functions of the 654
keV decay is $+6.2(7)\%$ for the central selection and
$-15.9(8)\%$ for the wing selection, assuming a pure M2
transition.

The R(t) functions of the 207 keV and the 654 keV $\gamma$-rays
de-exciting the $^{61m}Fe$ isomer exhibit an opposite phase and
the ratio between their amplitudes is 1.43(16). Using realistic
GEANT simulations \cite{geant} and assuming that the 654 keV and
207 keV transitions have pure M2 and M1 multipolarities,
respectively, and the level sequence is $9/2^+$$\rightarrow$
$5/2^-$$\rightarrow$ $3/2^-$, we estimated this ratio to be
1.30(6). A spatial distribution close to the fragmented beam spot
of the emitting $^{61}Fe$ isomers is assumed as well in order to
take into account the geometrical factor of the detection setup.
The good agreement between the two ratios (experimental and
simulated) indicates that the multipolarities and sequence of the
$\gamma$ transitions used for the determination of the
experimental alignment are correct.

The measured half-life of the $^{61}Fe$ isomer, $T_{1/2}=245(5)$
ns, is in good agreement with the previous measurements
\cite{grzywacz98}. For the g factor, a value of $-0.229(2)$ was
extracted by a $\chi ^2$ fit of the R(t) functions using
expression (\ref{fo:Rt}) with both frequency and initial phase,
$\alpha$, depending explicitly on $g$. The error on the fitted g
factor includes the errors on the effective field. The fitted R(t)
functions were constructed using the 654 keV direct decay of the
isomeric state for the central and wing selection of the fragments
momentum, respectively (fig. \ref{fig:rt_fits}).

\begin{figure}[t]
\includegraphics[scale=0.5]{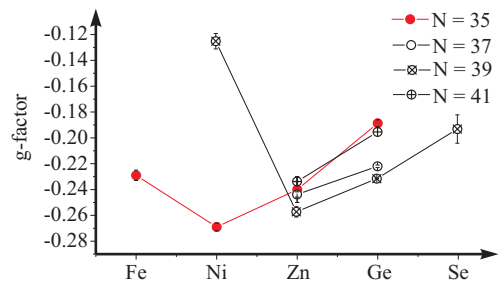}
  \caption{g factor systematics around N=40 for known $9/2^{+}$
isomeric states. The data are taken from
\cite{georgiev02,nucmomtab,muller89}. The unusual value for the g
factor of $^{67}Ni$ could be explained by proton excitations
across Z=28.} \label{fig:syst}
\end{figure}

In Fig. \ref{fig:syst} we present the measured g factors of
$9/2^{+}$ isomeric states around $N=40$, including also the g
factor of $^{61m}Fe$. The comparison with the other g factors of
known $9/2^{+}$ isomeric states in the region strongly supports
the $9/2^{+}$ spin and parity assignment for the $^{61m}Fe$
isomer. One can observe the symmetry with respect to $Z=28$ of the
g factors for $N=35$ chain and the increase of g factor values
when one goes away from $Z=28$, indicating an increase of core
polarization effects.

In order to extract detailed information about the valence orbital
occupation for the isomeric state, we have performed large scale
shell model (LSSM) and mean-field calculations using the
Hartree-Fock-Bogoliubov (HFB) formalism with an effective D1S
Gogny nucleon-nucleon force  \cite{decharge80}.

The shell model calculations were performed using the ANTOINE code
\cite{antoine} of the Strasbourg group in a valence space composed
of {\it f,p} and {\it $g_{9/2}$} active orbitals and having a
closed $^{48}Ca$ core. The interaction used is described in
\cite{sor02_prl88} and in the references therein.

The calculated free g factor is $g(9/2^{+})=-0.277$. The effective
value is $g(9/2^{+})=-0.1828$ if a quenching factor of 0.7 is used
for the nucleon spin g factor. The value of the quenching factor
is quite arbitrary because at present there is no systematical
comparison between experimental data and shell
model calculations into the considered space. Recently, it was
shown that for the {\it sd} and  {\it pf} shells, the
configuration mixing within the shells is enough to fully account
for the observed magnetic moment, without the use of the quenching
factor needed for Gamow-Teller beta decay \cite{honma04}. The
$9/2^{+}$ state is calculated to be at 720 keV and its wave
function is a mixture of a large number of configurations, but the
mean occupation of the $\nu g _{9/2}$ orbital is $\approx 1$. The
good agreement between the calculated and the experimental value
of the g factor indicates that the isomer is a $9/2^{+}$ state,
generated by a neutron in the $\nu g _{9/2}$ orbital whilst its
very mixed wave function could be related to a deformed potential.

\begin{figure}[t]
\includegraphics[scale=0.45]{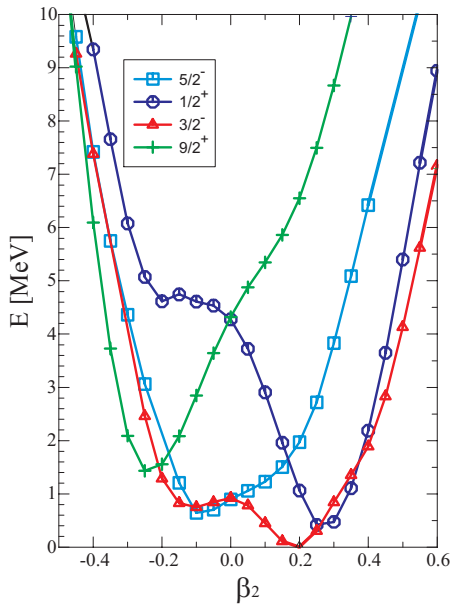}
  \caption{Potential energy curves of the ground state and of the
first three excited states of $^{61}Fe$ as functions of the
deformation parameter, $\beta_2$.} \label{fig:hfb}
\end{figure}

For a better understanding of the variation of the single particle
orbitals near the Fermi surface as a function of the nuclear
deformation, we have calculated the ground-state and excited
states within the Hartree-Fock-Bogoliubov (HFB) formalism. The
calculation beyond the mean-field approximation was performed
using the Generator Coordinate Method (GCM) with the Gaussian
Overlap Approximation \cite{libert99}. The ground state is found
to be a $K^\pi$ = $3/2^-$ prolate deformed state, $\beta_2=0.187$,
whereas an excited state $K^\pi$ = $9/2^+$ is predicted with an
excitation energy $E^*$ = 1640 keV and an oblate deformation,
$\beta_2=-0.227$.
\noindent Fig. \ref{fig:hfb} presents the HFB energies of the
states located near the Fermi level in $^{61}Fe$ as a function of
the deformation parameter, $\beta_2$.

The free g factor of the $J^\pi= 9/2^+$ state built from the
$K^\pi= 9/2^+$ HFB state is found to be $g({9/2^+})=-0.3477$
(without a quenching factor) and it remains practically unchanged
after applying the GCM method.

The comparison between the experimental value and the theoretical
predictions, LSSM and HFB, indicates a deformed potential for the
$9/2^+$ isomeric state. One can notice that the two extreme HF
orbitals ($9/2^+$ and $1/2^+$) generating from the spherical $\nu
g _{9/2}$ orbital (fig. \ref{fig:hfb}) are close to the Fermi
surface at moderate deformations but with opposite sign. In the
one particle plus rotor model \cite{stephens72}, this
particularity provides two possible ways of creating a
$J^{\pi}=9/2^+$ state, depending on how the unpaired neutron is
coupled to the rotating $^{60}Fe$ core. In the strong coupling
scheme \cite{nilsson_book}, the unpaired neutron spin couples to
the deformation of the core and the $9/2^+$ state has the
projection on the symmetry axis $K=9/2$. The second possibility is
that the unpaired neutron spin couples to the rotation of the core
and the total spin projection on the symmetry axis is degenerated
between $K=1/2$ and $K=-1/2$ due to the Coriolis interaction, so K
is no longer a good quantum number. Instead, the projection on the
rotation axis becomes a good quantum number and the spin of the
physical state can be $9/2^+$. We have then calculated the low
energy states in the frame of this model for $^{61}Fe$. Taking the
deformations indicated by the HFB calculations, for the two cases
presented above, the lowest state having a positive parity has a
$9/2$ spin and an excitation energy of about 850 keV. The g
factors are similar, which was predictable because the g factor
should not be sensitive to the deformation.

In conclusion, the measured g factor is in very good agreement
with the assigned $9/2^+$ spin and parity. From the comparison
with LSSM and HFB calculations there are indications that this
state is characterized by a deformed potential, induced by a steep
lowering of the Nilsson orbitals emerging from the spherical
${\nu}g_{9/2}$ neutron orbital as a function of deformation
parameter, $\beta_2$. In order to understand the coupling of the
unpaired particle to the $^{60}Fe$ core, it is important to
measure the sign of the quadrupole moment of the isomeric state,
requiring a spin polarized isomeric beam \cite{dafni84,hass84}.

The appreciably large residual alignment measured for the
$^{61}Fe$ and $^{54}Fe$ fragments indicates that fragmentation
reactions at intermediate energies can provide a powerful tool to
align ensembles of nuclear isomers of yet more exotic nuclei, thus
facilitating the determination of EM moments in neutron rich
nuclei and allowing the investigation of nuclear structure away
from stability.

We are grateful for the technical support received from the staff
of the GANIL facility. This work has been partially supported by
the Access to Large Scale Facility program under the TMR program
of the EU, under contract nr. HPRI-CT-1999-00019, the INTAS
project nr. 00-0463 and the IUAP project P5/07 of the Belgian
Science Policy Office. We are grateful to the IN2P3/EPSRC
French/UK loan pool for providing the Ge detectors. The Weizmann
group has been supported by the Israel Science Foundation. G.N.
and D.B. acknowledge the FWO-Vlaanderen for financial support.


\end{document}